\documentclass[conference]{IEEEtran}
\IEEEoverridecommandlockouts
\usepackage{cite}
\usepackage{amsmath,amssymb,amsfonts}
\usepackage{algorithmic}
\usepackage{braket}
\usepackage{graphicx}
\usepackage{subcaption}
\usepackage{multirow}
\usepackage{textcomp}

\usepackage{xcolor}
\usepackage{natbib}
\usepackage{booktabs}
\usepackage{amsmath}
\usepackage{hyperref}
\usepackage{tikz} \usetikzlibrary{shapes.geometric, arrows.meta, positioning}

\def\BibTeX{{\rm B\kern-.05em{\sc i\kern-.025em b}\kern-.08em
    T\kern-.1667em\lower.7ex\hbox{E}\kern-.125emX}}

\makeatletter 
\newcommand{\linebreakand}{%
  \end{@IEEEauthorhalign}
  \hfill\mbox{}\par
  \mbox{}\hfill\begin{@IEEEauthorhalign}
}
\makeatother

\begin{document}

\title{Quantum Optimization of the Color-of-Money Problem on IBM Quantum Hardware: A Proof-of-Concept Study\\
}

\author{\IEEEauthorblockN{Rahul Rana, Suman Kumar Roy,  M Girish Chandra}
\IEEEauthorblockA{\textit{TCS Research, Tata Consultency Services Limited, India} \\
\{rahul.rana2, suman.r2, m.gchandra\}@tcs.com}

}



\maketitle

\begin{abstract}

The Color-of-Money (CoM) problem is a financial optimization challenge in which enterprises must determine how to allocate and route funds across multiple accounts, currencies, and time periods while satisfying regulatory, operational, and liquidity constraints. As the number of assets, obligations, and planning periods increases, the decision space grows rapidly, making large-scale optimization increasingly challenging for classical approaches. In this work, we present a proof-of-concept quantum implementation of the CoM problem using IBM Quantum hardware. The original optimization problem is formulated as a constrained Mixed-Integer Linear Programming (MILP) problem and transformed into a Quadratic Unconstrained Binary Optimization (QUBO) representation suitable for quantum execution. We employ a Variational Quantum Algorithm (VQA)  with Pauli Correlation Encoding (PCE), a qubit efficient encoding technique and apply error-mitigation technique to improve solution fidelity on noisy quantum hardware. A prototype instance consisting of 3 assets, 3 obligations, and 2 time periods was executed on real IBM quantum hardware and benchmarked against classical exact solver SCIP. The quantum approach successfully reproduced the same optimal solution obtained classically, validating the correctness and feasibility of the proposed methodology. While current noisy quantum hardware does not outperform classical optimization on the evaluated small instance, the results demonstrate the feasibility of executing treasury optimization workloads on current quantum hardware and provide a foundation for future investigations.

\end{abstract}

\begin{IEEEkeywords}
Bank Treasury, Color of Money, Invoice Allocation, Liquidity Transfer, Quantum Finance, QUBO, Pauli Correlation Encoding, Quantum Optimization, Error Mitigation.
\end{IEEEkeywords}

\section{Introduction}
Efficient liquidity allocation is a fundamental challenge in enterprise treasury management and banking operations. Organizations must determine how available funds should be utilized to satisfy financial obligations while complying with operational policies, liquidity constraints, and regulatory requirements. In practice, money is often subject to restrictions regarding how and where it can be used, a concept commonly referred to as the Color of Money (CoM) \cite{bib1}. We formulate this problem as an Invoice Funding and Liquidity Transfer Problem, in which invoices must be assigned to eligible funding nodes and supporting liquidity transfers coordinated across a financial network. The problem incorporates funding eligibility, account-balance conservation, invoice deadlines, and time-dependent funding decisions, resulting in a constrained combinatorial optimization problem. Similar liquidity and collateral management problems have been identified as critical optimization challenges within the financial services industry, where increasing regulatory and operational complexity significantly impacts treasury decision-making. 

The CoM problem becomes increasingly difficult as the numbers of assets, obligations, and planning periods grow. Every additional funding source, invoice, or time period introduces new combinations of possible funding and transfer decisions, leading to rapid growth in the search space. Classical optimization methods such as Mixed Integer Linear Programming (MILP) can provide exact solutions for small and medium-sized instances; however, large-scale financial optimization problems often become computationally intensive due to the combinatorial nature of binary decision variables and operational constraints. Similar scalability challenges have been reported across a range of financial optimization applications, including portfolio allocation, collateral management, and liquidity optimization \cite{bib2}. 

Recent advances in quantum computing have motivated the exploration of quantum algorithms for combinatorial optimization problems. In particular, optimization models that can be expressed as Quadratic Unconstrained Binary Optimization (QUBO) \cite{bib3} problems are natural candidates for quantum optimization techniques. Variational quantum algorithm (VQA) \cite{bib4} have attracted considerable attention because they can be executed on current noisy intermediate-scale quantum (NISQ) hardware while leveraging a classical optimization loop to update quantum circuit parameters. Quantum and quantum-inspired optimization approaches have already been investigated for several financial applications, including portfolio optimization, currency arbitrage optimization, collateral optimization, and resource allocation problems, demonstrating the feasibility of representing practical financial decision-making tasks as QUBO models \cite{bib5}\cite{bib6}.

Most existing work in quantum finance has focused on portfolio optimization, risk management, and collateral allocation, whereas treasury-oriented CoM optimization problems have received limited attention in the quantum computing literature. Furthermore, relatively few studies evaluate optimization workflows on real quantum hardware and systematically compare the obtained solutions with exact classical benchmarks. Since current quantum devices are affected by noise and limited qubit counts, such proof-of-concept studies are essential for understanding the practical applicability of quantum optimization techniques and identifying suitable problem formulations for near-term quantum systems. 

In this work, we present a quantum optimization framework for the CoM problem. The original treasury optimization model is first formulated as a constrained MILP problem and subsequently transformed into a QUBO representation. We employ a VQA with Pauli Correlation Encoding (PCE) \cite{bib7} to solve the resulting QUBO on quantum system. To improve solution fidelity, Zero-Noise Extrapolation (ZNE) is applied \cite{bib8}. A proof-of-concept instance consisting of three assets, three obligations, and two time periods is benchmarked against classical exact solver Solving Constraint Integer Programs (SCIP) \cite{bib9}. The goal of this study is not to demonstrate quantum advantage, but rather to establish the correctness, and feasibility characteristics of a quantum approach for a real-world treasury optimization problem. 


\section{Methodology}
\label{Methodology}

\subsection{Mathematical Model}
Let $I$, $N$, $T$ and $E \subseteq N \times N$ denote set of invoices, funding nodes, periods, and permissible transfer edges respectively with parameter $D_i$ and $A_i$ are the due date and amount of invoice $i$ respectively. $elig_{i,n}\in \{0,1\}$ is CoM eligibility; $c_{n,m}$, and $\tau_{n,m}$ are the transfer cost, and delays to complete the payment from node $n$ to $m$; while $\tilde{c}_{i,n,t}$ refers the invoice payment cost and $b_{n,0}$ is the initial balance at node $n$. Binary variables $x_{i,n,t}$ indicates invoice $i$ is funded by node $n$ at time $t$, $f_{n,m,t}  \in \mathbb{R^+}$,  is the amount transfer from node $n$ to $m$ at time $t$, $p_{i,n,t} \in \mathbb{R^+}$ is the amount paid for invoice $i$ by node $n$ at time $t$ and $b_{n,t}$ is the balance of node $n$ at time $t$. The objective is to minimize the total cost.
\begin{equation}
    \label{eq:1}
    min ~Obj:\sum_{(n,m) \in E} \sum_{t \in T} c_{n,m} f_{n,m,t} + \sum_{i \in I}\sum_{n \in N}\sum_{t \in T} \tilde{c}_{i,n,t}x_{i,n,t}
\end{equation}
Subject to:
\begin{equation}
    \label{eq:22}
    b_{n,t} = b_{n,t-1}+\sum_m f_{n,m,t-\tau_{n,m}} - \sum_m f_{n,m,t} - \sum_i p_{i,n,t}, \:\: \forall n, t
\end{equation}

\begin{equation}
    \label{eq:3}
    \sum_{n \in N} \sum_{t \leq D_i} p_{i,n,t} = A_i,\:\: \forall\; i
\end{equation}

\begin{equation}
    \label{eq:44}
    p_{i,n,t} \leq A_i x_{i,n,t},\:\: \forall\; i,n,t
\end{equation}

\begin{equation}
    \label{eq:5}
    x_{i,n,t} \leq elig_{i,n},\:\: \forall\; i,n,t
\end{equation}

\begin{equation}
    \label{eq:6}
    \sum_{n \in N} \sum_{t \leq D_i} x_{i,n,t} = 1,\:\: \forall\; i
\end{equation}

Equation (\ref{eq:22}) denotes the flow balance constraint, equation (\ref{eq:3}) shows invoice completion. Equation (\ref{eq:44}) refers to the linking constraint between $x_{i,n,t}$ and $p_{i,n,t}$, whereas equation (\ref{eq:5}) denotes the eligibility criteria and equation (\ref{eq:6}) shows the invoice assignments before due date.

\subsection{QUBO Formulation}
The QUBO formulation of the above mathematical model is: 
\begin{equation}
    \label{eq:qubo}
    H_{QUBO} = Obj+\lambda_1H_1+\lambda_2H_2+\lambda_3H_3+\lambda_4H_4
\end{equation}

where penalty terms corresponding to the constraints in eq.(\ref{eq:22}), eq.(\ref{eq:3}), eq.(\ref{eq:44}), and eq.(\ref{eq:6}) are denoted by $H_1$,$H_2$, $H_3$ and $H_4$ respectively and are defined as follows:


\begin{gather*}
H_1 = \sum_{n\in N}\sum_{t=1}^{T} \Bigg( b_{n,t}-b_{n,t-1}  -\sum_m f_{n,m,t-\tau_{n,m}}  \nonumber\\  +\sum_m f_{n,m,t}  +\sum_i p_{i,n,t} \Bigg)^2 \end{gather*}

\begin{equation*}
    H_2 = \sum_{i\in I} \left( \sum_{n \in N} \sum_{t\le D_i} p_{i,n,t} -A_i \right)^2
\end{equation*}

\begin{equation*}
    H_3 = \sum_{i\in I} \sum_{n \in N} \sum_{t\in T} \left( p_{i,n,t}-A_i x_{i,n,t} + s \right)^2
\end{equation*}

\begin{equation*}
    H_4 = \sum_{i\in I} \left( \sum_{n \in N} \sum_{t\le D_i} x_{i,n,t} -1 \right)^2
\end{equation*}


    
$\lambda_1, \lambda_2, \lambda_3, \lambda_4$ are the respective penalty values. We have used  Verma and Lewis Method (VLM)\ \cite{bib10} to find the suitable values of  $\lambda_1, \lambda_2, \lambda_3 \ \text{and}\ \lambda_4$. Constraint eq.(\ref{eq:5}) is enforced directly while defining the $x_{i,n,t}$ variables.
The continuous variables and the slack variable $s$ are binary discretised as follows 

$p_{i,n,t} = \sum_{l=0}^{L}2^l pb_{l,i,n,t}$, $f_{n,m,t} = \sum_{l=0}^{L}2^l fb_{l,n,m,t}$, $b_{n,t} = \sum_{l=0}^{L}2^l bb_{l,n,t}$, and $s =\sum_{l=0}^{L}2^l sb_{l} $  and  $pb_{l,i,n,t}$, $fb_{l,n,m,t}$, $bb_{l,n,t}$, $sb_{l}$ are binary variables.

\subsection{Pauli Correlation Encoding}

A fundamental limitation of conventional gate-based quantum optimization algorithms, particularly the Quantum Approximate Optimization Algorithm (QAOA), is their direct one-to-one correspondence between binary decision variables and physical qubits. Consequently, a QUBO problem containing $(\eta)$ binary variables generally requires $(\eta)$ qubits, while the associated circuit depth increases with problem complexity and interaction density \cite{bib11}. Such scaling rapidly exceeds the computational capabilities of contemporary NISQ hardware, thereby restricting the applicability of standard QAOA formulations to relatively small optimization instances. To address this challenge, PCE provides a compressed representation in which multiple classical variables are embedded into correlation observables defined over a substantially smaller quantum register. By exploiting the combinatorial growth of pairwise pauli correlations, PCE reduces the qubit requirement from $(\mathcal{O}(\eta))$ to $(\mathcal{O}(\sqrt{\eta}))$ for two-body encodings, enabling significantly larger optimization problems to be represented within the constraints of current quantum devices~\cite{bib7}. 

In the PCE framework, each binary decision variable $(x_i \in \{0,1\})$ is encoded into the expectation value of a two-qubit pauli correlation observable 
\begin{equation} 
\label{eq:7} 
P_i=\sigma_{\alpha}^{(q)} \otimes \sigma_{\alpha}^{(r)}, \qquad \sigma_{\alpha}\in\{X,Y,Z\}, 
\end{equation} 
acting non-trivially on a selected qubit pair $(q,r)$ and as the identity on all remaining qubits. Only same-axis pauli correlations are employed, while mixed-axis observables are excluded. This design enabling simultaneous measurement of all encoded variables using a constant number of measurement bases. Such a property is critical for reducing measurement overhead and achieving hardware-efficient expectation-value estimation on NISQ platforms. To obtain a continuous and noise-resilient surrogate representation of binary variables, the measured pauli correlations are transformed through a nonlinear activation function, \begin{equation} 
\label{eq:8} 
\tilde{x}_i = \tanh \left(\alpha \langle P_i \rangle \right), \end{equation}
where $(\alpha > 0)$ is a scaling parameter controlling the sharpness of the encoding. The hyperbolic tangent maps correlation values smoothly into the interval $[-1,1]$, attenuates the effects of finite-sampling fluctuations, and progressively approaches a signed binary representation as $\alpha$ increases. This continuous relaxation improves the robustness of the optimization process in the presence of shot noise and hardware imperfections while preserving the discrete structure of the underlying combinatorial problem. 

\textbf{Variational Quantum Optimization Framework: } Given the penalty-augmented QUBO objective 
\begin{equation} 
\label{eq:9} 
H_{\text{objective}} = \sum_i a_i x_i + \sum_{i<j} b_{ij} x_i x_j, \end{equation} where $a_i,b_{ij}\in\mathbb{R}$ denote the linear and quadratic QUBO coefficients respectively. The PCE representation induces a fully differentiable variational objective expressed exclusively in terms of encoded pauli correlations, 
\begin{equation} 
\label{eq:4} 
\mathcal{L}_{\text{PCE}}(\theta) = \sum_i a_i \tilde{x}_i(\theta) + \sum_{i<j} b_{ij} \tilde{x}_i(\theta)\tilde{x}_j(\theta), \end{equation} where ~$ \tilde{x}_i(\theta) = \tanh\left( \alpha \langle P_i \rangle_\theta \right).$ The expectation values are evaluated with respect to the parameterized quantum state \begin{equation} |\psi(\theta)\rangle = U(\theta)|0\rangle, \end{equation} where \(U(\theta)\) denotes a variational quantum circuit parameterized by the trainable parameters \(\theta\). Optimization proceeds through a hybrid quantum-classical workflow in which the pauli correlation observables are estimated using the quantum Estimator primitive, while the circuit parameters are iteratively updated by a classical optimizer capable of handling noisy objective landscapes. Feasibility constraints are enforced implicitly through the penalty terms incorporated within the QUBO formulation, eliminating the need for explicit constraint handling during variational optimization. Following convergence, the continuous PCE representation is projected back onto the binary solution space through a thresholding operation, 
\begin{equation} \label{eq:binary_recovery} x_i^{*} = \frac{ 1+\operatorname{sgn} \left( \langle P_i\rangle_{\theta^{*}} \right) }{2}, \end{equation} where \(\theta^{*}\) denotes the optimized circuit parameters.

\subsection{Overall Architecture}
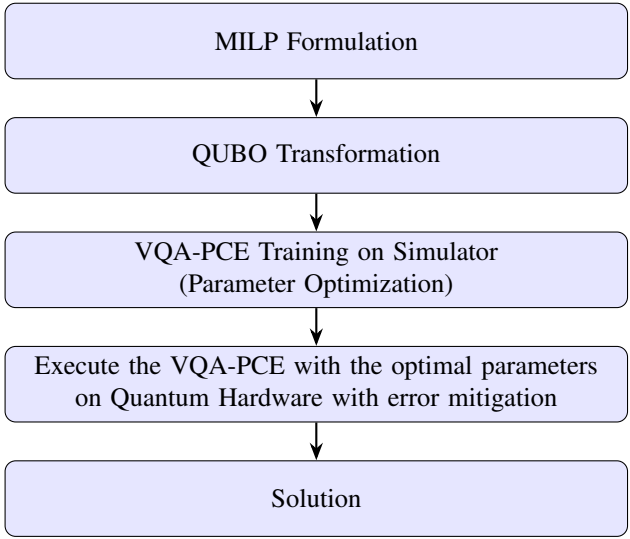
\begin{figure}[t] 
\centering
\begin{tikzpicture}[ node distance=0.5cm, block/.style={ rectangle, rounded corners, draw=black, fill=blue!10, align=center, text width=8cm, minimum height=1cm }, arrow/.style={ thick, -{Stealth} } ] \node (milp) [block] {MILP Formulation}; \node (qubo) [block, below=of milp] {QUBO Transformation}; \node (sim) [block, below=of qubo] {VQA-PCE Training on Simulator\\(Parameter Optimization)}; \node (hw) [block, below=of sim] {Execute the VQA-PCE with the optimal parameters \\ on Quantum Hardware with error mitigation}; \node (sol) [block, below=of hw] {Solution}; \draw [arrow] (milp) -- (qubo); \draw [arrow] (qubo) -- (sim); \draw [arrow] (sim) -- (hw); \draw [arrow] (hw) -- (sol); \end{tikzpicture} \caption{Overall workflow of the quantum optimization framework.} \label{fig:workflow} \end{figure}

The proposed framework adopts an end-to-end hybrid quantum-classical optimization pipeline for solving the constrained combinatorial optimization problem. The overall architecture is illustrated in Fig.~\ref{fig:workflow} and consists of four sequential stages: (i) MILP model creation, (ii) QUBO transformation, (iii) simulator-driven variational training using PCE, and (iv) hardware execution with error mitigation. 

In the first stage a constrained MILP model is created based on the input data.
The constrained MILP model is then transformed into a QUBO representation by discretized continuous variables into binary variables and encoding all constraints through penalty functions. The resulting QUBO objective can be expressed as equation (\ref{eq:9}). This formulation enables direct compatibility with variational quantum optimization algorithms. 

Next, the QUBO has been encoded using the PCE framework. Unlike conventional QAOA-based approaches that require one qubit per decision variable, PCE maps optimization variables onto two-body pauli correlation observables, substantially reducing qubit requirements. A parameterized quantum circuit \(U(\theta)\) is constructed and executed initially on a noise-free quantum simulator. The simulator serves as a training environment for identifying high-quality variational parameters. 

The variational loss function is evaluated through pauli correlation measurements and minimized using a classical optimizer within a hybrid quantum-classical loop. Specifically, the simulator estimates the expectation values of the encoded pauli observables, which are transformed into continuous surrogate variables and used to compute the PCE objective. Iterative parameter updates continue until convergence, yielding an optimized parameter vector \(\theta^{*}\). 

After training convergence, the optimized variational parameters \(\theta^{*}\) are transferred directly to the target quantum hardware. Rather than performing full optimization on noisy quantum processors, only the final circuit evaluation is executed on hardware, significantly reducing quantum execution costs. To improve the reliability of measured expectation values, ZNE is incorporated during hardware execution. Multiple noise-scaled circuit instances are generated through circuit folding, and the corresponding expectation values are extrapolated toward the zero-noise limit. The error-mitigated expectation value of pauli observables are subsequently converted into binary decision variables using equation (\ref{eq:binary_recovery}). 

The resulting hybrid workflow combines the scalability advantages of PCE, the efficient parameter discovery enabled by simulator-based training, the enhanced hardware reliability provided by ZNE. Together, these components form a practical NISQ-compatible optimization framework capable of solving larger QUBO instances while maintaining solution quality under realistic quantum hardware constraints.



\section{Experimental Study}
\label{experiments}

\subsection{Experimental Setup}
The proposed quantum optimization framework was evaluated on a proof-of-concept CoM instance consisting of three assets, three invoices, and two time periods. The dataset was intentionally kept small to fit within the constraints of current noisy quantum hardware while preserving the key characteristics of the Invoice Funding and Liquidity Transfer Problem.
The classical benchmark was implemented using SCIP and executed on a workstation equipped with an AMD Ryzen 5 7535U processor (12 logical cores) and 16 GB of RAM. Classical runtimes reported in this work correspond to execution on this hardware platform. VQA with PCE is executed on $ibm$-$aachen$ quantum processor \cite{bib12}. 
\subsection{Results}
The primary objective of the experiment was to evaluate whether the proposed quantum formulation could reproduce the optimal solution obtained by exact classical optimization methods.
SCIP successfully identified the optimal funding plan with an objective value of $151$. As shown in Table \ref{tab:result}, the VQA-PCE implementation executed on IBM Quantum hardware recovered the same optimal objective value, demonstrating the correctness of the QUBO construction, quantum encoding, and optimization workflow.
Although the quantum implementation required a longer execution time than the classical solvers, it successfully identified the global optimum despite the presence of hardware noise and current NISQ-era limitations. The current experiment therefore demonstrates solution correctness and practical feasibility, rather than computational advantage.

\begin{table}[!h]
\centering
\caption{Classical and Quantum Optimization Results}
\label{tab:result}
\begin{tabular}{|l|c|l|l|}

\hline
\multicolumn{1}{|c|}{Method} & \begin{tabular}[c]{@{}c@{}}Objective \\ Value\end{tabular} & \multicolumn{1}{c|}{\begin{tabular}[c]{@{}c@{}}Optimal Solution\\  Recovered\end{tabular}} & \multicolumn{1}{c|}{\begin{tabular}[c]{@{}c@{}}Execution \\ Time (Sec)\end{tabular}} \\ \hline
SCIP                         & 151.00                                                       & Yes                                                                                        & 0.02                                                                                 \\ \hline
VQA-PCE    & 151.00                                                        & Yes                                                                                        & 31.00                                                                                   \\ \hline
\end{tabular}
\end{table}

\section{Conclusion and Future Works}
\label{conclusion}
This paper presented a proof-of-concept implementation of the CoM optimization problem on IBM Quantum hardware. The problem was formulated as a QUBO and solved using a VQA with PCE. For a test instance consisting of three assets, three obligations, and two time periods, the quantum implementation reproduced the same optimal solution obtained using exact classical solver. Experimental results demonstrate the correctness and feasibility of the proposed approach and show that meaningful treasury optimization problems can be executed on current quantum hardware using appropriate error-mitigation techniques. These findings establish an initial foundation for future research on quantum methods for financial and treasury optimization.

Future work will focus on extending the proposed framework to larger and more realistic CoM scenarios involving additional assets, obligations, currencies, and planning periods. We also plan to investigate more advanced optimization strategies, and error-mitigation techniques to further improve solution quality and execution efficiency. The proposed workflow provides a foundation for evaluating quantum approaches to increasingly complex treasury and financial optimization problems as quantum computing technology continues to advance.




\bibliographystyle{unsrt}
\bibliography{bibliography}


\end{document}